\documentstyle[12pt]{article}

\begin{document}

\centerline{\bf Proceedings of the Conference}

\bigskip

\centerline{\bf QUANTUM THEORY:}

\centerline{\bf RECONSIDERATION OF FOUNDATIONS}

\bigskip

\bigskip

\centerline{V\"axj\"o (Smaland), Sweden, 17-21 June, 2001}

\bigskip

\bigskip

\bigskip

\bigskip

\bigskip

\bigskip

\centerline{Editor Andrei Khrennikov}

\centerline{International Centre for Mathematical Modelling}

\centerline{in Physics , Engineering and Cognitive Science}

\centerline{V\"axj\"o University, Sweden}

\bigskip

\bigskip

\bigskip

\bigskip

\bigskip

\bigskip

\bigskip

\bigskip

\bigskip

\bigskip

\bigskip

\bigskip

\centerline{\bf V\"axj\"o University Press}

\centerline{Series: Mathematical Modelling in Physics,}

\centerline{ Engineering and Cognitive Science, v.2}

\newpage
\newpage
\bigskip

{\bf Citations:}

N. Bohr:

{\small Notwithstanding the difficulties which are involved in the 
formulation of quantum theory it seems that its essence may be expressed
in the so called quantum postulate, which attributes to any atomic process an 
essential discontinuity, or
rather individuality, completely foreign to classical physics and symbolized by the 
Planck's quantum action.}

\medskip

P. Dirac:

{\small An act of observation is thus necessarily accompanied by some disturbance of the object observed... If a
system is small, we cannot observe it without producing a serious disturbance and, hence we cannot expect 
to find  any causal connection between the results of our observation.}

\medskip

R. Feynman:

{\small But far more fundamental was the discovery that in nature the laws of combining
probabilities were not those of the classical probability of Laplace.} 

\medskip

A. Peres:

{\small What we call quantum paradoxes are quantum phenomena that lead
to paradoxes if we try to interpret them in a classical way.} 

\medskip

S. Gudder:

{\small Where does the Hilbert space $H$ come from?
Why does the probability have the postulated form?
Why does a physical theory which must give real-valued results
involve a complex amplitude state?
Why must a quantum particle exhibit wave behaviour (wave-particle dualism)?
Must quantum mechanics be nonrealistic (a quantum system only has properties when 
they are observed)?
Is there a realistic solution of the EPR problem ?} 

\medskip

A. Plotnitsky: 

{\small Bohr's interpretation of quantum mechanics as complementarity may be seen,
or interpreted, in informational terms.  The main reason for this view is
that this interpretation is grounded in the epistemologically radical
assumption that no physical properties of quantum objects or of processes
involving them are describable by means of quantum theory or, it appears,
any theory, and still more radically that no such properties are ascribable
to them.  The formalism of quantum theory and its physical interpretation
are seen as referring to certain, statistically predictable, effects of the
(quantum) interaction between quantum objects and the measuring instruments
(partially described in terms of classical physics) upon the latter. }

\newpage

\section{Preface}

\bigskip

This volume\footnote{To order this volume please contact Kerstin.Broden@adm.vxu.se} constitutes the proceedings of the Conference "Quantum Theory: Reconsideration 
of Foundations" held in V\"axj\"o (Smaland, Sweden), 17-21 June, 2001.

The organizing committee of the conference: C. Fuchs (Bell's Laboratory, USA), A. Khrennikov (V\"axj\"o University, Sweden), 
P. Lahti (Turku University, Finland).

The purpose of the conference (the fourth in the series of V\"axj\"o conferences) 
was to bring together scientists (physicists, mathematicians and philosophers) who are interested in foundations
of quantum physics. An emphasis was made on both theory and experiment, the underlying objective being to offer to 
the physical, mathematical and philosophic communities a truly interdisciplinary conference as a privileged
place for a scientific interaction.
Due to the actual increased role of foundations in the development of quantum information theory as
well as the necessity to reconsider foundations at the beginning of the new millennium, the organizers 
of the conference decided that it was just the right time for taking the scientific risk of trying this.

Quantum theory was created at the beginning of XXth century. For many years
it is considered as the a well established scientific discipline with its own philosophy, logic, 
methodology, probability, geometry and having numerous experimental confirmations. Moreover, quantum 
theory has a lot of applications: physics of elementary particles, nuclear weapon and energy and the last years 
also - {\it quantum information, computing, cryptography and teleportation.} It must be underlined that 
the latter applications of quantum theory (quantum information and so on)
are essentially stronger related to foundations
than, for example, nuclear physics. My contacts with Soviet nuclear
physicists demonstrated that they were merely interested in computational apparatus of 
quantum theory. An interpretation of these computations was not yet a subject of deep investigations.
But quantum information, computing, cryptography are deeply connected with many fundamental problems in
foundations of the quantum theory. It seems that it is no longer the case that we have
restrict ourself just to the use of computational methods of quantum mechanics. 

Some fundamental problems are still unsolved (despite enormous efforts  
from physicists, mathematicians, philosophers). I think that there are following main reasons for such a  situation 
in foundations of quantum theory:

\medskip

(A) Great complexity of problems under considerations.

(B) Some wrong pathways that were chosen in the 1920s and `cemented' by 
the authority of the fathers - creators of the quantum theory.

(C) Insufficient knowledge in mathematics by physicists and insufficient
knowledge in physics by mathematicians working in quantum physics.

One of the main problems with wrong pathways
is the huge overestimation of the authority of the fathers - creators of the quantum 
theory. At the moment it is really impossible (or at least very hard) to discuss consciously 
fundamental problems such as e. g. presented  in Gudder's list of questions, see citations.
The standard arguments are citations of Bohr, Heisenberg and Pauli or their adherents 
as well as pointing out that there are great experimental confirmations of quantum theory
and, finally, (especially last years) recalling Bell's story.

First we pay attention to the experimental confirmation argument. 
Yes, quantum computations give right predictions 
that are confirmed by experiments. But we have to split quantum computation 
machinery and interpretations of results of such computations as
well  as functioning of quantum computation apparatus. 

One of the main purposes of the present conference was to discuss 
the views of the fathers - creators to foundations of quantum mechanics. We were not oriented 
to criticize ideas of the fathers - creators. The main attitude was to understand better their
ideas via deep studying of their works. In fact, some wrong pathways as well as prejudices were 
not consequences of e.g. Bohr's original views, but a rather vague
understanding of these views. I think that philosophers H. Folse 
and A. Plotnitsky in their talks made great contributions to the correct understanding of Bohr's views.
Many things that are rigidly associated with Bohr's name were, 
in fact, never directly 
presented  by Bohr. Even so called orthodox Copenhagen interpretation of quantum mechanics
was not originally formulated by N. Bohr:
{\it `The wave function provides the complete description of an {\it individual} quantum system'.}

We now pay attention to the (C)-source of
the unsatisfactory situation in 
foundations, namely the insufficient level 
of exchange of ideas between physicists, mathematicians and philosophers. 
It is in our power to improve (C) - 
to organize a series of meetings oriented to the collaboration of physicists (theoreticians as well as experimenters), 
mathematicians and philosophers.  The present conference, "Quantum Theory: Reconsideration of Foundations",
was the fourth meeting with such an aim that  has been taken place in V\"axj\"o the last three years:
{\it "Quantum days in V\"axj\"o "}, November 1999: $p$-adic numbers and space-time, 
Bell's inequality and foundations of probability theory, quantum information;
{\it "Bohmian Mechanics - 2000"}, May 2000:
computer simulation of Bohmian trajectories, 
noncommutative geometry and Bohmian mechanics, 
typically and probability, Bohmian model for mental processes;
"Foundations of Probability and Physics", November-2000: the
role of probability in EPR-Bell considerations, probability 
and information, $p$-adic probability and $p$-adic reality, Kolmogorov complexity, 
von Mises' probability theory and quantum mechanics.

The role of probability in foundations of quantum mechanics was one of the 
most important problems discussed during the last V\"axj\"o conference. 
I would like to pay attention to an extended discussion on the possibility to use Bayesian (subjective) 
probability theory in quantum information induced by talks of C. Fuchs and R. Schack, see
the fundamental paper of C. Fuchs in this volume.

This discussion played an important role in the clarifying of the probabilistic structure of quantum
information theory. It also attracted the attention to the role of a mathematical model of probability 
theory in quantum formalism. In particular, I presented some reasons in the favour of the 
frequency (von Mises) probabilistic model. The `Orthodox Bayesian approach' of Fuchs--Schack 
was criticized from the frequency point of view. There are no doubts that we can use Bayesian method 
of statistical hypotheses and, moreover, it is very convenient in quantum information theory. However, it
is very doubtful that quantum probabilities can be introduced as a measure of our personal belief. Well, 
it may be belief, but belief based on frequency information. On the other hand, C. Fuchs and I. Pitowsky 
presented strong critical arguments against von Mises' frequency probability theory (including 
the impossibility to verify statements depending on $N \to \infty$ number of trials). 

In many talks 
and during the round table it was discussed the role of probability in the EPR-Bell framework: L. Accardi - theory 
of chameleons and analogy between Bell's inequality and inequalities for the sum of angles in a triangle 
for various geometric models,  W. De Baere - fluctuations of hidden variables, A. Khrennikov - 
non-Kolmogorov models of probability theory, e.g. frequency (von Mises) or $p$-adic 
and Bell's inequality, I. Pitowsky - 
Bell's inequality as an example of an inequality for random variables derived 
by J. Boole in XXth century,
generalizations of Bell's inequality. Investigations on the probabilistic structure 
of Bell's assumptions performed e.g.
by L. Accardi,  W. De Baere, S. Gudder, A. Khrennikov, W. De Muynck, I Pitowsky 
are the good illustrations to the (C)-problem in the study of foundations of quantum mechanics. In 
fact, by taking into account very important mathematical assumption - the use of Kolmogorov's 
probability model by J. Bell,
we can easily see that Bell's inequality does not contradict to {\it local realism.} The above
authors presented various models in that Bell's inequality is violated.
However, all these investigations are considered by the majority of the quantum community as pure 
mathematical, non-relevant to real physics.

The crucial role of a mathematical probabilistic model in the foundations of 
quantum mechanics, namely quantum interference, was discussed in the talk of A. Khrennikov. 
It was demonstrated that `wavelike' {\it{interference of probabilities of alternatives}} can be derived without
using wave arguments. We must only leave the domain of applications of the conventional probabilistic model, 
Kolmogorov - 1933, and use the frequency or contextual models of probability theory, see the paper in this volume.

One of the most important problems, 
intensively discussed during the conference, was the problem of an {\it interpretation} of 
a wave function. It was quite surprising that the orthodox 
Copenhagen interpretation was denied by the majority of participants. 
Personally I (as well as W. De Baere) supported a {\it contextual statistical realist interpretation}, `V\"axj\"o interpretation', 
by that quantum randomness is context (=complex of experimental conditions) depending randomness.
Such an interpretation (in the opposite to Bohr's informationalcontextualism, see citation of 
A. Plotnitsky and further considerations)
does not imply the impossibility to create finer description of physical reality 
(if you like hidden variables) than given by the quantum theory, see my text in this volume.
The {\it individual realistic interpretation} used in operational quantum physics
was strongly supported by P.  Lahti during the round
table of the Conference.

During the round table, the large group of participants, J. Bub, C. Fuchs,
D. Mermin, A. Plotnitsky,..., supported various forms of
{\it information-oriented interpretation} of quantum mechanics. To some degree,
such interpretations may be seen as following Bohr's interpretation (or a
certain type of interpretation of Bohr interpretation).  According to this
type of view, in a rough outline, quantum mechanics does not refer to
objective properties of physical objects themselves under investigation, but
deals with the relationships between and predictions concerning
informational quantities.

The orthodox Copenhagen interpretation was strongly supported by
J. Summhammer. In some sense J. Summhammer presented the general 
point of view of experimenters working in neutron interferometry. Here neutron is
imagined as a wavelike object that is split into two or more pathways with further i
interference of corresponding parts. 
Regarding to interpretations of a wave functions discussed during the conference, 
we should also notice the presentation of the {\it{Bohmian interpretation}} 
by B. Hiley. In Hiley's presentation Bohmian mechanics was merely an 
attempt to testify how long we can proceed in quantum theory by using 
classical mechanical formalism.
Finally, we remark that {\it{many worlds interpretation}} 
did not induce strong enthusiasm among the participants (despite very enthusiastic 
propaganda of this interpretation by L. Vaidman).

The round table discussion on interpretations was closely related to the discussion on foundations 
of {\it {quantum computing.}} Especially interesting problem was related to interpretations 
of {\it{quantum parallelism.}} It is really the hard problem. In majority of works on quantum
computing there are really claimed that all values of a function under the computation are really 
calculated parallel. Regarding to quantum parallelism, it is important the remark of R. Jozsa. 
He pointed out that parallel computation of all values is merely a convenient 
mathematical picture for this quantum process. He also presented very important 
ideas on the role of entanglement of quantum computation and considered some possible 
interpretations.

Quantum information considerations demonstrated that there is a danger that manipulating 
with pure qubits we can forget real physics. In this way there might be produced results 
that are valid for pure qubits, but might not be applied to real quantum systems. I. Volovich 
underlined that such a problem we have already in Bohm-Bell framework for the EPR experiment. 
He presented strong arguments that by taking into account quantum mechanical processes in 
space-time we have to modify standard Bell's inequality.

I would like to recall the words of Russian academician Krylov: 

\centerline{\it {"Mathematics is a kind of mill. It mills all that we put in it."}}

It seems that quantum formalism is nothing than a quantum mathematical mill.
The only difference from other mathematical mills that are used in other domains 
of physics, e.g. Newtonian mill, is that we do not know well what we put into the quantum mill. 
We see the result of working of the quantum mill - very good mill that can be used for many purposes. 
But we cannot  `see' so called elementary particles, `quantum grain',
without to change their features (and, as a consequence, features of quantum mill at the output). 
Such a situation induces various prejudices. One of the strongest ones is the identification of
some features of the quantum mathematical mill with physical features of quantum systems. 

I have the feeling that, in fact, quantum theory was based on two distinct discoveries. 
One of them was in the domain of physics - {\it discreteness} of energy as well as some other observables. 
The second one was the purely mathematical discovery, namely discovery of the calculus of context depending
probabilities (inducing interference rule for addition of probabilities)
and the possibility to represent contextual probabilistic calculus as Hilbert space probabilistic
calculus (see e.g. my paper in this volume).

In fact, contextual probabilistic calculus was only
occasionally discovered in the connection with investigations of elementary particles. In principle,
it might be discovered in the process of purely probabilistic investigations, e.g. in XVIII th or XIX th century. 
One of the problems was very preliminary stage of the 
development of the foundations of probability theory at the beginning of XXth century. In fact, 
the measure-theoretical approach to probability theory was developed at the same time when M. Planck and 
A. Einstein created foundations of quantum theory. If you read Einstein's papers, you see 
that he should work with probabilities by using intuitive arguments - real mathematical theory 
of probability was created by Kolmogorov 25 years later! But even Kolmogorov's
theory of probability did not provide mathematical tools sufficient for describing of quantum
statistical data. Kolmogorov's probability model was the {\it{fixed context model}}. 
And in quantum physics we need to use a context-variable model for probability theory. 

The absence of an adequate probabilistic theory induced the prejudice that quantum statistical 
data demonstrated extremely unusual (`nonclassical') features. Thus some features of the contextual
probabilistic mill were interpreted as features of elementary particles. Moreover, quantum physicists, 
M. Born, W. Heisenberg, P. Dirac, did (rather unconsciously) the great mathematical discovery.
They found that transformations of context depending probabilities, interference of probabilities,
$$
P=P_1+ P_2+ 2\sqrt{P_1 P_2} \cos \theta
$$
can be represented 
as linear transformations in a Hilbert space. So, instead of manipulating with nonlinear transformations of probabilities, 
we can work by using linear algebra. This Hilbert space probabilistic mill  was not separated
from, `quantum grain', elementary particles. Some special features of the Hilbert space mill 
were considered as features of elementary particles. 

In fact, an attempt to separate the quantum mill from quantum 
grain also  was done by L. Hardy who derived quantum theory from five very reasonable axioms,
see his paper in this volume. We underline that such a derivation might be in principle performed in XIXth century, 
far before quantum experimental discoveries. 

The conference and the present volume give the good example of the fruitful collaboration 
between physicists, mathematicians and philosophers. We would like to thank the Swedish 
Science Foundation and V\"axj\"o University (through Rector's "Strategic Investigations Foundation") 
for financial support. We would also like to thank Prof. Magnus S\"oderstr\"om, the Rector of
V\"axj\"o University, for the support of the fundamental investigations, in particular, 
for his enormous efforts to create the "Mathematical Modeling in Physics, Engineering and Cognitive Sciences" 
specialization of V\"axj\"o University.

\medskip

Andrei Khrennikov

\medskip

Director of International Center for Mathematical Modeling 

in Physics, Engineering and Cognitive Sciences

\newpage

\newpage

{\bf  CONTENTS}

\bigskip

Preface \hspace*{\fill} 5\\

\medskip 

1. The EPR correlations and the chameleon effect  \hspace*{\fill} 15\\

{\it L. Accardi,  M. Regoli}

\medskip 

2. Application of $p$-adic analysis to models 

of breaking of replica symmetry \hspace*{\fill} 31\\

{\it V. A. Avetisov, A. H. Bikulov, S. V.Kozyrev}

\medskip 

3. Bohmian mechanics for stock market  \hspace*{\fill} 41\\

{\it O. Choustova}

\medskip 

4. Subquantum nonreproducibility and the complete 

local description of physical reality  \hspace*{\fill} 59\\

{\it W. De Baere, W. Struyve}

\medskip 

5. Holonomic quantum logic gates  \hspace*{\fill} 75\\

{\it M. Ericsson}

\medskip 

6. Bohr's conception of the quantum mechanical state of a system

and its role in the framework of complementarity  \hspace*{\fill} 83\\

{\it H. J. Folse}

\medskip 

7. The anti-V\"axj\"o interpretation of quantum mechanics  \hspace*{\fill} 99\\

{\it C. A. Fuchs}

\medskip

8. Quantum theory from intuitively reasonable axioms  \hspace*{\fill} 117\\

{\it L. Hardy}

\medskip 

9. Set theory to physics  \hspace*{\fill} 131\\

{\it J. Helmick}

\newpage

\medskip 

10. From the Heisenberg picture to Bohm: a new perspective
on 

active information and its relation to Shannon information  \hspace*{\fill} 141\\

{\it B. J. Hiley}

\medskip 

11. On foundations of quantum theory  \hspace*{\fill} 163\\

{\it A. Khrennikov}

\medskip 

12. Quantum-like formalism for cognitive measurements  \hspace*{\fill} 197\\

{\it A. Khrennikov}

\medskip 

13. Two slits interference is compatible with 

particles' trajectories  \hspace*{\fill} 215\\

{V. V. Kisil}

\medskip 

14. An approximation sequence for phase observables  \hspace*{\fill} 227\\

{\it P. Lahti and J.-P. Pellonp\"a\"a}

\medskip 

15. The Kochen-Specker paradox in a probabilistic setting  \hspace*{\fill} 233\\
 
{\it J.-A. Larsson} 
 
\medskip 

16. Fuzzy space-time and phase space geometry 

as approach to quantization  \hspace*{\fill} 245\\

{\it S. N. Mayburov}

\medskip 

17. Whose Knowledge?  \hspace*{\fill} 261\\

{\it N. D. Mermin}

\medskip 

18. Bound states in quantum tubes  \hspace*{\fill} 271\\

{\it B. Nilsson}

\medskip 

19. Unspeakable quantum information  \hspace*{\fill} 283\\

{\it A. Peres and P. F. Scudo}

\medskip 

\newpage

20. Range theorems for quantum probability and entanglement  \hspace*{\fill} 299\\

{\it I. Pitowsky}

\medskip 

21. Quantum atomicity and quantum information: Bohr, Heisenberg,

and quantum mechanics as an information theory  \hspace*{\fill} 309\\

{\it A. Plotnitsky}

\medskip 

22. Pancharatnam revisited  \hspace*{\fill} 343\\

{\it E. Sj\"{o}qvist}

\medskip 

23. Comments on some recently proposed 
experiments that should 

distinguish Bohmian mechanics from 
quantum mechanics  \hspace*{\fill} 355\\

{\it W. Struyve, W. De Baere}

\medskip 

24. Conservation of statistical information and quantum laws  \hspace*{\fill} 367\\

{\it J. Summhammer}

\medskip 

25. Dialogue on classical and quantum between 

mathematician  and experimenter  \hspace*{\fill} 385\\

{\it J. Summhammer, A. Khrennikov}

\medskip 

26. Extension of quantum information theory 

to curved spacetimes  \hspace*{\fill} 397\\

{\it D. R. Terno}

\medskip

27. Probability and many worlds interpretation of quantum theory  \hspace*{\fill} 407\\

{\it L. Vaidman}

\medskip 

28. Towards information theory in space and time  \hspace*{\fill} 423 \\

{\it I. V. Volovich}

\medskip 

29. Discrete time leads to quantum-like interference of 

deterministic particles  \hspace*{\fill} 441\\

{\it Ya. Volovich, A. Khrennikov}

\medskip

30. Interference effect for probability distributions of 

deterministic particles  \hspace*{\fill} 455\\

{\it Ya. Volovich, A. Khrennikov}

\medskip

31. Quantum mechanics as quantum information 

(and only a little more)  \hspace*{\fill} 463\\

{\it C. A. Fuchs}

\end{document}